\documentclass[aps,prb,reprint,showpacs,showkeys,superscriptaddress,preprintnumbers,amsmath,amssymb,longbibliography]{revtex4-2}
\usepackage[utf8]{inputenc}
\usepackage[utf8]{inputenc}
\usepackage{multirow}
\usepackage{graphicx}
\usepackage{bm}
\usepackage{mathtools}
\usepackage{dcolumn}
\usepackage{picture}
\usepackage{physics}
\usepackage[colorlinks=true,urlcolor=blue,linkcolor=blue,citecolor=blue]{hyperref}
\usepackage{etoolbox}
\usepackage{natbib}
\usepackage{amsmath}
\usepackage{times}
\usepackage{color}
\usepackage{xcolor}
\usepackage{soul}
\usepackage[per-mode=symbol]{siunitx}
\usepackage{amssymb}
\usepackage{amsmath,latexsym}
\usepackage[normalem]{ulem}
\usepackage{url}

\newcommand{\Yb}[1]{YbI$_3$}
\newcommand{\J}[2]{$\rm J_{eff}$}

\begin{document}
	\title{Atypical antiferromagnetic ordering in single crystalline quasi-2D honeycomb magnet YbI$_3$}
	
	\author{Nashra Pistawala}
	\affiliation{Department of Physics, Indian Institute of Science Education and Research, Pune 411008, Maharashtra, India}
	
	\author{Luminita Harnagea}
	\affiliation{I-HUB Quantum Technology Foundation, Indian Institute of Science Education and Research, Pune 411008, India}

         \author{Sitaram Ramakrishnan}
	\affiliation{Department of Physics, Indian Institute of Science Education and Research, Pune 411008, Maharashtra, India}
    \affiliation{I-HUB Quantum Technology Foundation, Indian Institute of Science Education and Research, Pune 411008, India}

	\author{Priyanshi Tiwari}
	\affiliation{UGC-DAE Consortium for Scientific Research, University Campus, Khandwa Road, Indore 452001, Madhya Pradesh, India}

    \author{M. P. Saravanan}
	\affiliation{UGC-DAE Consortium for Scientific Research, University Campus, Khandwa Road, Indore 452001, Madhya Pradesh, India}
 
	\author{Rajeev Rawat}  
	\affiliation{UGC-DAE Consortium for Scientific Research, University Campus, Khandwa Road, Indore 452001, Madhya Pradesh, India}

	\author{Surjeet Singh}
	\email[email:]{surjeet.singh@iiserpune.ac.in}
	\affiliation{Department of Physics, Indian Institute of Science Education and Research, Pune 411008, Maharashtra, India}
	
	\date{\today}
	
	\begin{abstract}
		
		%A rare earth-based $\rm J_{eff} = 1/2$ honeycomb magnet \Yb~serves as the closest realization to Kitaev quantum spin candidate $\alpha-$RuCl$_3_$. 
  Here, we study YbI$_3$, a quasi-2D layered material with Yb atoms arranged on an ideal honeycomb network of edge-sharing YbI$_6$ octahedra, analogous to the low-temperature phase of $\alpha-$RuCl$_3$. High quality single crystals of YbI$_3$ are grown from Yb and I as starting precursors, using the vapor transport technique. %Their crystal structure is found to be trigonal ($R\Bar{3}$) from  single crystal x-ray diffraction. %The Raman spectrum at room-temperature shows six distinct Raman active phonon modes in agreement with the point group symmetry%centered at 38.4, 56.4, 69.4, 94.0, 118.2, 142.5 cm$^{-1}$
  The grown crystals are characterized by single crystal x-ray diffraction, Raman spectroscopy, magnetization, and heat capacity probes. The %low-temperature magnetic and thermodynamic behavior of \Yb~ suggests that the 
  crystal-field split ground state of Yb$^{3+}$ in \Yb~ is a well-isolated Kramers doublet with an effective moment $\rm J_{eff} = 1/2$. Upon cooling, the low-temperature heat capacity of \Yb~ reveals a broad peak at $\rm T_1 = 0.95$~K due to short-range ordering of the Yb moments, followed by a sharp peak at $\rm T_2 = T_N = 0.6$~K due to long-range ordering. %, and an upturn upon cooling below 0.25~K (nuclear)
  %measurement reveals the onset of short-range magnetic ordering at 0.95~K and the long-range antiferromagnetic ordering at 0.6~K. 
   %The magnetic entropy obtained after subtracting the Nuclear Schottky tail approaches 82\% of Rln2 at 4~K. 
  The magnetic behavior is found to be weakly anisotropic with $\chi^\parallel > \chi^\perp$, where $\chi^\parallel$ and $\chi^\perp$ refers to the in-plane ($H \parallel ab$) and out-of-plane ($H \perp ab$) susceptibilities. The 2~K isothermal magnetization saturates at $\rm \approx~1.5~\mu_B/Yb^{3+}$ (in-plane) and $\rm \approx~1~\mu_B/Yb^{3+}$ (out-of-plane), suggesting the anisotropy to be easy-plane type.  %anisotropic with %the Land\'{e} \textit{g}-factor $g^\parallel$ and $g^\perp$, respectively, 2.3 and 2.7, where $\parallel$ and $\perp$ refers to the orientations with respect to the \textit{c}-axis, aligned perpendicular to the honeycomb layer. %, obtained from the magnetic susceptibility analysis indicates suggesting a nearly Heisenberg-like nature of the Yb$^{3+}$ spins. 
  The %critical exponent $\alpha$ obtained from the low-temperature power-law behavior of 
  low-temperature heat capacity, well below T$_N$, is found to vary as T$^\alpha$ with $\alpha~\approx~2.5$, %differing from the ideal 3D Heisenberg antiferromagnet, 
  indicating a possible unconventional magnetic ground state for YbI$_3$.
		
	\end{abstract}

	\maketitle
	\label{Introduction}
 
	\section{Introduction}
	\label{Intro}
 
        Quantum magnetism is one of the frontier areas in condensed matter physics, %. The quantum magnets exhibit 
   showcasing unusual ground states, such as, spin liquids with emergent quasiparticles, and exotic spin textures driven by competing interactions~\cite{Savary_2017,balents2010spin}. %The magnetic frustration refers to a situation where competing interactions prevent the spins from aligning. %Quantum spin liquid is an exotic state of matter characterized by collective behaviors that cannot be described solely in terms of individual particles. Instead, it involves emergent quasiparticles, leading to novel magnetic ground states~\cite{Savary_2017,balents2010spin}. %
   The %exotic quantum states in these systems 
   magnetic interactions in these systems are generally frustrated, which prevents %them from achieving a 
   simple, ordered spin arrangements, %. The absence of long-range ordering down to the lowest possible temperatures or 
   %The ground states that emerge in such scenarios %the occurrence of fractionalized moments, part frozen/part dynamic as in some compounds based on the pyrohclore structure, leads to a 
   leading to complex or unconventional ground states. The %geometrically frustrated lattices such as the 
   spin systems based on triangular, Kagome, or pyrochlore lattices are model geometrically frustrated systems, where the competing interactions result from the lattice geometry which does not allow all the nearest-neighbor couplings to be simultaneously satisfied. %lattice following Anderson's proposal of the resonating valence bond model~\cite{ANDERSON1973153}. 
   
   The Kitaev model, introduced by Alexei Kitaev in the year 2006~\cite{KITAEV20062}, opened-up a new frontier for exploring magnetic frustration. This groundbreaking theoretical framework describes a spin-1/2 system on a two-dimensional honeycomb lattice, where the bond-dependent nearest-neighbor couplings give rise to a strong frustration. The ground state of Kitaev model is a novel quantum spin liquid with Majorana fermions as quasiparticle excitations. 
   
   The quest for experimental realization of Kitaev model gained momentum following the seminal work by Jackeli and Khaliullin~\cite{PhysRevLett.102.017205}, who laid out a roadmap for uncovering potential Kitaev candidates among the spin-orbit-assisted Mott insulators, specifically the $4d^5$ ruthenates and $5d^5$ iridates. In these compounds, the complex interplay between the strong spin-orbit coupling, octahedral crystal electric field splitting, and electronic correlations give rise to a $\rm J_{eff} = 1/2$ Kramers doublet ground state with highly anisotropic bond-dependent interactions, which serves as a prerequisite for realizing the Kitaev model. 
   
   The electronic and magnetic properties of candidate materials, including,  Na$_2$IrO$_3$~\cite{PhysRevB.82.064412,PhysRevB.83.220403,PhysRevB.85.180403}, $\alpha$- Li$_2$IrO$_3$~\cite{PhysRevLett.108.127203,PhysRevB.93.195158}, $\beta$-Li$_2$IrO$_3$~\cite{PhysRevLett.114.077202,ruiz2017correlated}, $\gamma$-Li$_2$IrO$_3$~\cite{PhysRevLett.113.197201}, H$_3$LiIr$_2$O$_6$~\cite{kitagawa2018spin}, Cu$_2$IrO$_3$~\cite{doi:10.1021/jacs.7b06911}, Cu$_3$LiIr$_2$O$_6$~\cite{C6DT00798H}, Ag$_3$LiIr$_2$O$_6$~\cite{TODOROVA20111112},$\alpha$-RuCl$_3$~\cite{PhysRevB.90.041112,PhysRevB.92.235119,nasu2016fermionic,PhysRevLett.114.147201,PhysRevB.93.134423,doi:10.1126/science.aah6015,bruin2022robustness} has been extensively studied in recent years to explore the manifestations of Kitaev physics.
   
   The $f$-electron systems, based on the rare-earths %on the other hand, characterized by stronger spin-orbit coupling compared to 4d/5d systems, 
   with $\rm J_{eff} = 1/2$ Kramer's doublet ground state and edge-shared octahedra %ligand environments surrounding the central rare-earth ion, 
   have also been theoretically predicted to host Kitaev physics~\cite{PhysRevB.95.085132,PhysRevB.98.054408,PhysRevB.99.241106,Motome_2020}. However, suitable materials embodying this concept remain sparse. Xing \textit{et al}. recently investigated the crystal growth and magnetic properties of YbCl$_3$, aiming to explore its potential for Kitaev physics~\cite{PhysRevB.102.014427}. YbCl$_3$ is a two-dimensional van der Waals material, where the Yb atoms form a slightly distorted honeycomb net, %. Within the honeycomb layers, 
   where each Yb atom has nearest-neighbors at 3.886~\AA~and 3.864~\AA~\cite{PhysRevB.100.180406}. The previous studies reveal short-range ordering of Yb$^{3+}$ moments ($\rm J_{eff} = 1/2$) around 1.2 K and a Néel-type antiferromagnetic long-range order at 0.6 K, with Yb spins antiferromagnetically aligned within each honeycomb plane, pointing primarily along the \textit{a}-axis with a small tilt of $\sim~5^\circ$ along the \textit{c}-axis~\cite{PhysRevB.102.014427, hao2021field, sala2021van}. %The inelastic neutron scattering measurements suggest that YbCl$_3$ manifests as an ideal two-dimensional honeycomb system with Heisenberg interactions~\cite{sala2023field}, devoid of frustrations or anisotropic interactions. Beyond a single magnon mode, they observed signatures of a two-magnon continuum marked by sharp Van Hove singularities~\cite{sala2021van}. 
   
    A recent study by Matsumoto \textit{et al}. provides evidence for Bose-Einstein condensation in YbCl$_3$. When the magnetic field is aligned parallel to the \textit{a}-axis, YbCl$_3$ has been argued to undergo a transition from Heisenberg to XY-like state, ultimately achieving a field-polarized state at $\rm H_s = 5.93$~T, marked as the quantum critical point. Upon reducing the field slightly below $H_s$, YbCl$_3$ allegedly behaves as a two-dimensional gas of bosons, showing signatures of Bose-Einstein condensation possibly stabilized by a small but non zero interlayer interaction ($\Lambda_{\perp}\approx$ 0.1~mK and $\Lambda_{\parallel}\approx$ 5~K, where $\Lambda_{\perp}$ and $\Lambda_{\parallel}$ are interlayer and intralayer nearest neighbor interactions)~\cite{matsumoto2024quantum}. Hence, YbCl$_3$ is posited as an exemplary two-dimensional system for studying quantum magnetism on a honeycomb lattice. 
    
    Conversely, YbBr$_3$, also comprising $\rm J_{eff} = 1/2$ Heisenberg spins on a honeycomb lattice, demonstrates no long-range ordering down to 100~mK. Based on their inelastic neutron scattering data, Wessler \textit{et al}. argued that despite lacking Kitaev-type anisotropic interactions, YbBr$_3$ exhibits pronounced quantum fluctuations, %on a honeycomb lattice, thereby introducing a novel perspective for 
    perhaps signature of a quantum spin liquid ground state~\cite{wessler2020observation}. Considering the novel magnetic behavior associated with the YbCl$_3$ and YbBr$_3$, we set out to investigate the previously unexplored member YbI$_3$ of this family.
   
    Here, we report the crystal growth of a new two-dimensional van der Waals material, YbI$_3$, using the vapor transport technique. %The high quality of the crystals is verified using Laue diffraction and X-ray diffraction on the ab plane of single crystals. The chemical composition was verified using FESEM-EDX, giving Yb:I ratio close to 1:3. 
    YbI$_3$ crystallizes in $R\Bar{3}$ symmetry, determined using the single crystal x-ray diffraction. % technique and verified using Raman spectroscopy. 
    In this structure, the Yb$^{3+}$ ions form an ideal honeycomb network, with equidistant nearest neighbors with the honeycomb plane. This structure is analogous to the low-temperature (T~$<$~150~K) phase of $\alpha$-RuCl$_3$, %Thus, the $C_3$ rotational symmetry in the \textit{ab}-plane, 
    suggesting that YbI$_3$ is a possible candidate material for studying Kitaev physics. %in the 4\textit{f} electron systems. 
    
    %The Raman spectrum at 300~K shows six distinct modes. 
    The ground state of Yb$^{3+}$ in \Yb~ is shown to be a well-isolated Kramers doublet with an effective moment, $\rm J_{eff} = 1/2$. The magnetic behavior is found to be slightly anisotropic. While the in-plane saturation moment at 2~K is $\rm \approx 1.5~\mu_B$/Yb$^{3+}$, it is $\rm \approx 1~\mu_B$/Yb$^{3+}$ for the out-of-plane orientation, suggesting an easy-plane behavior analogous to YbCl$_3$ (here in-plane and out-of-plane refers to the orientation of the applied field with respect to the \textit{ab}-plane). %and this difference becomes more pronounced upon cooling. 
    The low-temperature heat capacity %of YbI$_3$ 
    shows %three distinct features: 
    a broad peak at T$_1$ = 0.95 K due to short-range ordering of the Yb moments, %spin interactions, characteristics for low dimensional systems, 
    and a sharp peak at T$_2$ = 0.6 K that marks the onset of long-range antiferromagnetic ordering. 
    %and an upturn upon cooling below 0.25~K. %The magnetic entropy, calculated after subtracting the low-temperature upturn arising due to nuclear Schottky, which shows entropy release approaches to Rln2, 
    %The entropy consideration suggests a Kramers doublet ground state; however the recovered entropy remains almost 20\% short of the Rln2 value up to 4~K, suggesting the spread of short-range correlations up to much higher temperatures, characteristics for low dimensional systems.%The Curie - Weiss temperatures obtained from the magnetic susceptibility measurement down to 2 K are negative for $H\parallel c$ and $H\perp c$ orientation, indicating the antiferromagnetic interactions between the spins along two orientations. 
    %The \textit{g}-value does not show any significant difference along the two orientations, 
    %The magnetic behavior is found to be isotropic, suggesting the Heisenberg nature of the Yb spins. 
    The value of critical exponent $\alpha$ obtained from fitting the low-temperature heat capacity using the power-law is 2.5, different from $\alpha=3$ for a 3D Heisenberg antiferromagnets or $\alpha=2$ for the Heisenberg spins on a 2D lattices, suggesting a possible unconventional quantum ground state for this system.  
	
 \section{Experimental Methods}
	\label{Exp}
    Single crystals of \Yb~ are grown using the vapor transport technique in a two-zone furnace. The details of the crystal growth are given in sec. \ref{CG}. The x-ray diffraction patterns %measurements on the ab plane of the crystals 
   are recorded using a Bruker D8 advance diffractometer in the Bragg Brentano geometry employing Cu K$_\alpha$ ($\lambda = 1.5406$ {\AA}) radiation. The $2\Theta$ values ranged from 5$^\circ$ to 90$^\circ$ with a step size 0.02$^\circ$. For this purpose, a thin plate-like single crystal was placed over a oven-dried glass-slide and completely covered by a layer of Kapton tape to avoid degradation due to moisture. 
   
   Single crystal x-ray diffraction (SCXRD) was carried out using a Bruker Smart Apex Duo diffractometer employing Mo-K$_\alpha$ radiation ($\lambda = 0.70173$ {\AA}) at 250~K and~150 K using an open-flow nitrogen cryostat. A tiny crystal of approximately 1.630~$\mu$m lateral dimensions was covered in paratone oil to avoid exposure to the atmosphere and mounted on a goniometer loop. The intensities were integrated using the Crystalis Pro~\cite{Crystalispro}. The refinement was performed using Jana 2006~\cite{PetříčekDušekPalatinus+2014+345+352}. 

  % \indent The crystals were oriented using a Laue camera (Photonic Science, UK) in back scattering geometry using a tungsten source with wavelength $\rm \lambda = 0.35{\AA} - 2.5{\AA}$, accelerating voltage 30~kV, tube current 0.3~mA. The Laue pattern was analyzed using Orient Express 3.4 (V 3.3) software package.

   The layered morphology of the single crystals was confirmed using a scanning electron microscope (SEM) (FEI Make), and the Chemical composition was verified using an Energy Dispersive x-ray (EDX) (Quanta 2003D) detector attached to the SEM instrument in the Secondary Electron (SE) mode. The EDX spectra was collected at $15-20$ different points, spread across the platelet-shaped single crystal specimen (see the Supplementary Material~\cite{supplemental}), to obtain the averaged elemental composition. The elemental mapping was also performed over a micron size area of the single crystal to check the compositional homogeneity of the grown crystal.

   Raman spectra were collected at room temperature in the back scattering configuration using a Horiba Jobin-Yvon LabRAM HR spectrometer equipped with liquid nitrogen cooled Charge-Coupled Detector (CCD) and a laser of 532 nm as the source of excitation. The excitation was maintained at 75\% of the maximum power, and the accumulation time for each spectrum was 10 s with 30 iterations each time to improve the resolution and intensity of the Raman modes. For the purpose of obtaining Raman spectrum, the \Yb~ crystal was coated with a thin-layer of Apiezon N grease to prevent degradation due the measurement.

   Magnetic measurement between $2-300$~K are performed at the UGC-DAE CSR, Indore using the VSM probe in a Physics Property Measurement System (PPMS), Quantum design, USA. The heat capacity was measured $0.11$~K to $4$~K using a dilution insert in the PPMS at UGC-DAE-CSR, Indore. Due to highly hygroscopic nature of YbI$_3$, the crystal was covered with an unknown amount of Apiezon N grease, hence the addenda could not be measured accurately. However, in the temperature range of our measurement, the magnetic specific heat of YbI$_3$ is $3-4$ orders of magnitude higher than that of the Apiezon N grease, and can therefore be safely ignored while analyzing the measured data.     

	\section{Results and Discussion}
	\label{RD}
	\subsection{Crystal growth of YbI$_3$}
	\label{CG}
    
    \begin{figure}
		\centering
		\includegraphics[width= 0.9\columnwidth]{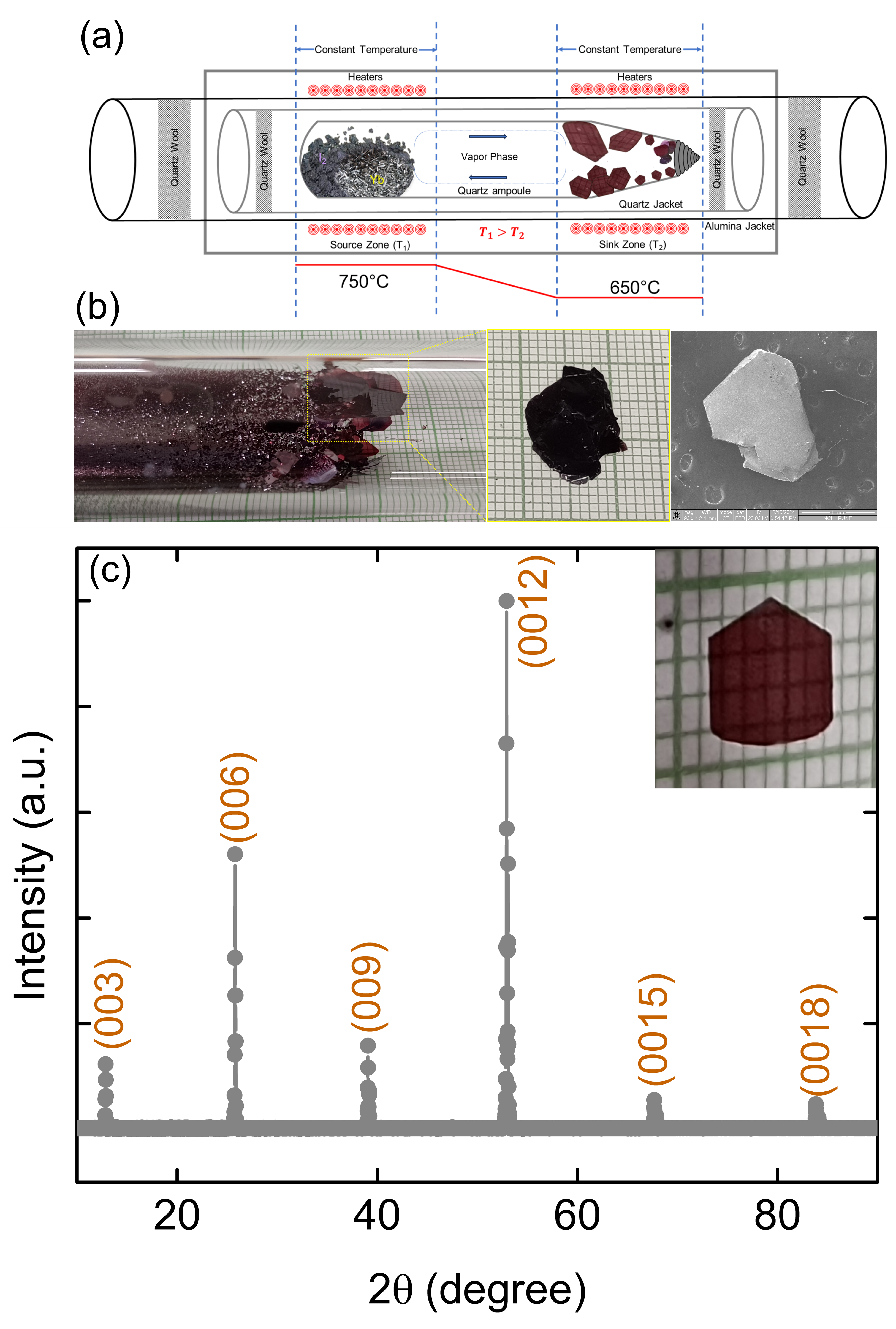}
		\caption {(a) Schematic of vapor transport reaction for \Yb~ in a sealed quartz ampoule kept in a temperature gradient furnace. (b) Actual growth ampoule (left), containing several small crystals, and a large cm-sized single crystal (middle).  A scanning electron microscope image of a small crystal specimen used for chemical analysis (right). (c) x-ray diffraction was conducted in Bragg Brentano geometry, displaying the (0 0 l) reflections which confirms the ab - plane of the single crystal specimen shown in the inset.} 
	\label{Fig_Image}
 \end{figure}
   
    High-quality single crystals of \Yb~ are grown using the physical vapor transport (self-transport) technique. Ytterbium powder (99.9\% REO, Alfa Aesar) and Iodine flakes (ACS reagent, $\geq$~99.8\%, Sigma Aldrich) are used in their elemental form as starting precursors. The stoichiometric quantities of the starting materials are loaded in a quartz ampoule, measuring approximately 15~cm in length, with an inner diameter of about 2~cm and wall-thickness of 2~mm, resulting in a volume of around $40 - 50$~cm$^3$. To compensate for iodine sublimation losses during the evacuation process, an additional 10\% excess of iodine is included. The entire procedure of handling and storing the precursors is carried in an argon-filled glove box with oxygen and moisture content $<$ 0.1 ppm. Before loading the precursors into the ampule, thorough cleaning steps are undertaken. The ampule undergoes sonication in acetone, followed by rinsing with deionized water. Subsequently, it is preheated to 1000$^\circ$C to eliminate residual moisture and organic impurities sticking to its inner walls. The filled ampule is evacuated down to $10^{-5}$ torr, and thereafter flame-sealed under dynamic vacuum. Given iodine's high vapor pressure at room temperature (approximately 100~Pa at 309~K), the ampule was kept under an ice bath during evacuation and sealing to minimize iodine losses. The sealed ampoule is then kept in a two-zone furnace under a temperature gradient $\Delta$T = 100$^\circ$C,  with the source temperature set at 750$^\circ$C ($\rm T_{source}$) and the sink temperature($\rm T_{sink}$) at 650$^\circ$C. The schematic of vapor transporter reaction is shown in Fig.~\ref{Fig_Image}(a). Optimal growth conditions for producing the largest and highest quality single crystals involve simultaneous cooling of both the zones at a rate of 0.3$^\circ$C~h$^{-1}$ down to $\rm T_{source} = 600^\circ C$ and $\rm T_{sink} = 400^\circ C$. Subsequently, both zones of the furnace are cooled simultaneously to room temperature. 
    
    The experiments performed using different growth conditions are summarized in Table S2. (See Supplemental Material)~\cite{supplemental}. Shiny, plate-like single crystals, with typical lateral dimensions of 1~cm $\times$ 1~cm, and dark-brown/transparent appearance, as shown in Fig.~\ref{Fig_Image}(b), are obtained. The plate-like morphology of the grown crystals is a reflection of the underlying layered or quasi-two-dimensional structure of \Yb~. A scanning electron microscope image of a small crystal specimen is shown in Fig.~\ref{Fig_Image}(b). The chemical composition, obtained using EDX, confirms the 1:3 ratio of Yb:I (see Fig. S1 and Table S2. of the Supplemental Material)~\cite{supplemental}. 
    
    The x-ray diffraction pattern collected from a plate-like single crystal specimen is shown in Fig.~\ref{Fig_Image}(c). The specimen used is shown in the inset. The pattern is collected in the Bragg Brentano geometry, revealed diffraction peaks with miller indices $(00l)$, where $l = 3, 6, ...$, suggesting that the large lateral surface of the measured specimen is oriented perpendicular to the \textit{c}-axis. %Polycrytsalline YbI$_3$ was first synthesized by Asprey et al. in the year  1964~\cite{asprey1964preparation}. They used Yb metal and iodine, sealing them in a quartz tube. This sealed ampoule was then encapsulated in a steel vessel with closed ends. The material was heated to 500°C over 8 h and then maintained at 500°C for 16 h. The iodine pressure at 500°C was approximately 30~atm. In the current work, the growth procedure uses only 2-3~atm of iodine pressure at 750°C, demonstrating that the higher iodine pressure reported earlier is not necessary for growing crystals of YbI$_3$.
    
    The crystals are found to be extremely hygroscopic, decomposing into white powder and eventually into a liquid droplet when exposed to the ambient atmosphere. Therefore, due care is required during sample preparation and while loading the sample for measurements in a cryostat as described in Sec.~\ref{Exp}.
    
    %and ensure that the crystal specimen remains shielded from the ambient atmosphere, even for brief periods.

	\subsection{Crystal Structure}
	\label{CS}
    %\label{CS}
        \begin{figure*}[hbt!]
		\centering
		\includegraphics[width=1.8\columnwidth]{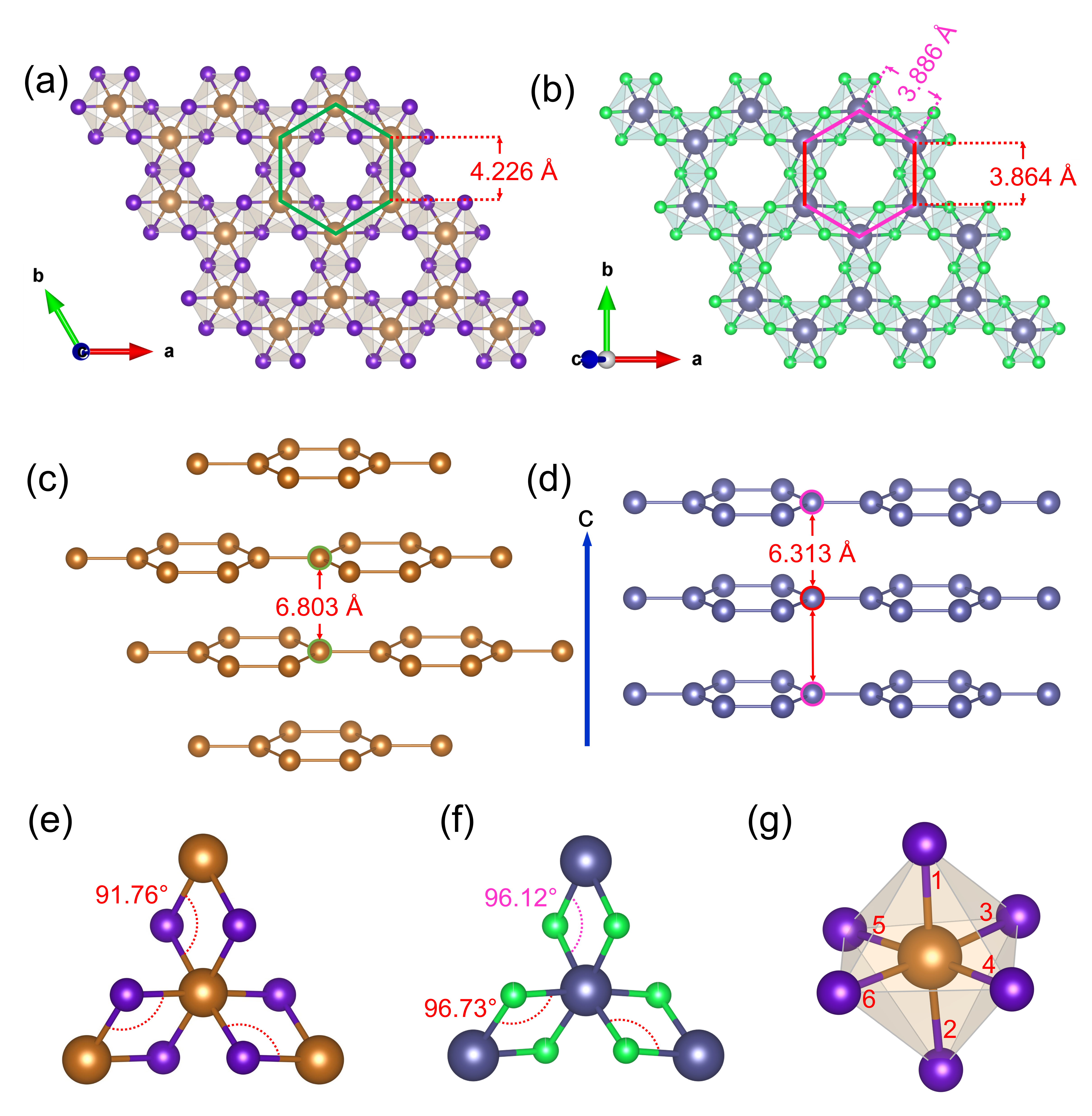}
		\caption {Crystal structure of \Yb~. (a) Projection of the \textit{ab}-plane showing the arrangement of Yb ions on a honeycomb lattice. The hexagonal rings formed by the Yb atoms, exemplified using green solid lines, is regular with each side 4.226(1)~{\AA}. (b) Analogous image for YbCl$_3$ is shown for comparison. In this case the hexagonal ring is slightly distorted with Yb-Yb distances 3.886~\AA~and 3.864~\AA~\cite{PhysRevB.100.180406}. (c) Stacking of the honeycomb layers of \Yb~ perpendicular to the \textit{c}-axis. The nearest neighbor Yb-Yb distance along the \textit{c}-axis is 6.803(4)~{\AA}. (d) Analogous image for YbCl$_3$ is shown for comparison, where the nearest neighbor Yb-Yb separation is 6.313~\AA~\cite{PhysRevB.100.180406}. (e) and (f) comparison of Yb-I-Yb bond angles along the three nearest neighbors in the \textit{ab}-plane. These angles are uniformly 91.76(3)$^\circ$ in YbI$_3$, and 96.12$^\circ$ and 96.73$^\circ$ in YbCl$_3$~\cite{PhysRevB.100.180406}. (g) A regular octahedron of I atoms around the Yb atom. Each Yb-I bond distance is close to 2.942(3)~{\AA}.}
		\label{Fig_CS}
	\end{figure*}
     
    The symmetry of YbI$_3$ was found to be trigonal $R\Bar{3}$ (\# 148) from single crystal x-ray diffraction, which is in agreement with the L Asprey et al.\cite{asprey1964preparation}. Temperature-dependent single crystal x-ray diffraction measurements at 250 K and 150 K show an absence of structural transition and that the symmetry remains trigonal as shown in Fig.~S2. (see Supplemental Material)~\cite{supplemental}, unlike $\rm \alpha-RuCl_3$, which exhibits a change of symmetry from room-temperature monoclinic ($C2/m$) to low-temperature trigonal ($R\Bar{3}$) symmetry below a temperature of 150~K~\cite{PhysRevB.91.094422} The symmetry of the lattice is further substantiated by the reciprocal space cuts shown %. Since we could not investigate the structure below 150 K, we have no knowledge about the structural phase transition below 150 K, if any.      
        
     Crystal structure of \Yb~ in the \textit{ab}-plane is shown in Fig.~\ref{Fig_CS}(a), illustrating the honeycomb network of Yb atoms similar to $\rm \alpha$-RuCl$_3$~\cite{do2017majorana} and YbBr$_3$~\cite{wessler2020observation}. Each Yb atom is coordinated to six neighboring I atoms in an almost ideal edge-sharing octahedral arrangement. In the honeycomb layer, the Yb-Yb bond distance is 4.226(1)~{\AA}, while the distance along the c-axis is 6.803(4){~\AA} as shown in Fig.~\ref{Fig_CS}(c), where the stacking of the honeycomb layers perpendicular to the \textit{c}-axis can be seen. The Yb-I-Yb bond angle in the honeycomb layer is 91.76(3)$^\circ$ (see panel (e) in Fig.~\ref{Fig_CS}). The octaheron of ligands I atoms around the central Yb atom is shown in Fig.~\ref{Fig_CS}(g). %The arrangement of \Yb~ layers along the crystallographic c-axis, where three layer stacking sequence similar to $\alpha-$ RuCl$_3$~\cite{park2024emergence} is clearly shown in Fig.~\ref{Fig_CS}(b) with a single interlayer nearest neighbor at a distance of 6.80 {\AA}. 

    The details of the data collection strategy, crystallographic information and refinement parameters are summarized in Table~\ref{SCXRD_CD}.  The atomic positions with their corresponding isotropic displacement parameters (ADPs) are listed in Table~\ref{Table_AP}. \par
     
     The crystal structure of \Yb~ is different from YbCl$_3$, where the later crystallizes with a monoclinic $C2/m$ crystal structure~\cite{PhysRevB.102.014427,PhysRevB.100.180406}, analogous to the room-temperature phase of $\alpha-$RuCl$_3$. However, unlike $\alpha-$RuCl$_3$, which undergoes a structural phase transition to $R\Bar{3}$ symmetry below 150~K, YbCl$_3$ remains in the $C2/m$ symmetry over the whole temperature range. In $C2/m$ symmetry, the hexagonal ring of the Yb atoms is not perfect with two different Yb-Yb bond distances of 3.886~{\AA} and 3.864~{\AA}. The Yb-Cl-Yb bond angles also take two different values, namely 96.12$^\circ$ and 96.73$^\circ$, thus lacking the $C_3$ symmetry~\cite{PhysRevB.100.180406}. %The change in the symmetry from $C2/m$ to $R\Bar{3}$ not only changes the stacking sequence of the honeycomb layers perpendicular to the \textit{c}-axis, but more importantly, the low-temperature phase exhibits a three-fold, $C_3$, rotational symmetry in the honeycomb plane, which is broken in the room temperature $C2/m$ phase, and hence lacking in YbCl$_3$.%,  thereby affecting the local symmetry that governs the magnetic interactions. 
    The effect of symmetry breaking has been discussed in the recent study by Kim et. al~\cite{PhysRevB.109.L140101}. The $C_3$ symmetry with 90$^\circ$ bond angles and larger inter-layer separation along the \textit{c}-axis are ideal criteria for realizing the Kitaev model~\cite{park2024emergence}, which minimizes the Heisenberg interaction, allowing Kitaev interactions to dominate, as shown in the case of $\alpha-$ RuCl$_3$~\cite{yadav2016kitaev,do2017majorana}. As mentioned above, YbCl$_3$ has a slightly distorted honeycomb net with different bond angles and with more number of nearest neighbors along the \textit{c}-axis~\cite{PhysRevB.100.180406}. YbBr$_3$, on the other hand, has a similar structure as that of YbI$_3$ and low-temperature phase of $\alpha-$ RuCl$_3$ but with small site-disorder on the Yb site~\cite{wessler2020observation}. To summarize, between the three Yb based halides compared here, YbI$_3$ appears to have the closest resemblance to the low-temperature phase of $\alpha-$ RuCl$_3$.

\begin{table}[!]
\caption{\label{SCXRD_CD}%
Crystallographic data and refinement parameters for SCXRD of \Yb~ single crystal at 250 K .}
\centering
\begin{ruledtabular}
\begin{tabular}{cc}
Temperature (K) &  250  \\
Crystal system & Trigonal  \\
Lattice system/Setting & Hexagonal  \\
Space group (Point group) & $R\Bar{3}$ ( $C_{3i}^{2}$)\\
 No.  & 148   \\
$a$ (\AA{}) & 7.3201(8)  \\
$c$ (\AA{}) & 20.4545(27)  \\
Volume (\AA{}$^3$) & 949.2(3)  \\
$Z$ & 6 \\
Wavelength (\AA{}) & 0.71073  \\
Detector distance (mm) & 40  \\
Scan type & Full sphere ($\phi$, $\omega$ scans) \\
Rotation per image (deg) & 0.8  \\
Exposure time (sec) & 10 \\
$(\sin(\Theta)/\lambda)_{max}$ (\AA{}$^{-1}$) & 0.743374 \\
Absorption, $\mu$ (mm$^{-1}$) & 29.261  \\
T$_{min}$, T$_{max}$ & 0.84, 1.08  \\
Criterion of observability & $I>3\sigma(I)$ \\
No. of  reflections measured & 10072  \\
Unique reflections (obs/all) & 561/728  \\
$R_{int}$  (obs/all) &0.1022/0.1054  \\
No. of parameters &14 \\
$R_{F }$   (obs) &0.0556 \\
$wR_{F }$ (all) &0.0720 \\
GoF (obs/all) &3.98/3.50 \\
$\Delta\rho_{min}$, $\Delta\rho_{max}$(e \AA$^{-3}$) &-3.59, 3.77  \\
\end{tabular}
\end{ruledtabular}
\end{table}

 \begin{table}[h]
 \caption{\label{Table_AP}%
Atomic Coordinates x, y, z and atomic displacement parameters(ADPs) of \Yb~ single crystal at 250 K in {\AA}$^2$ . Unique reflections (obs/all) = 561/728, criterion of observability: $I>3\sigma(I)$, $R_{F } = 0.0556$, No. of refined paramateres = 14, Refinement method used: least-squares on F, Space group: $R\Bar{3}$. The complete data including the anisotropic atomic displacement parameters, are shown in Table S3.(see Supplemental Material)~\cite{supplemental}.}
\centering
\begin{ruledtabular}
 \begin{tabular}{c c c c c c c}
  Atoms & Wyck & x & y & z & Occ. & $U_{iso}^{eqi}$\\[0.5ex]
 \hline\\[0.5ex]
  Yb & 6c & 0 & 0 & 0.3337(4) & 1 & 0.0182(3)\\
  I & 18f & 0.3229(1) & 0.3341(1) & 0.4165(1) & 1 & 0.0228(4)  \\
  \end{tabular}
  \end{ruledtabular}
   \end{table}

 \subsection{Raman Spectroscopy}
	 \label{RS}
      
       As discussed above, YbI$_3$ has a Trigonal symmetry with the point group symmetry $C_{3i}^{2}$. According to the group theory, the irreducible representation of the zone-center Raman active phonon modes in \Yb~ are $\Gamma(C_{3i}^{2}) = 4A_g+4^1E_g+4^2E_g$. %whereas for $D_{3d}$ point group, the representation of the phonon modes is $\Gamma(D_{3d})=4A_{1g}+2E_g$. 
       However, if the inter-layer coupling is treated as a small correction, the single-layer of YbI$_3$ has the $D_{3d}$ point group symmetry, rendering six Raman active modes, $\Gamma (D_{3d}) = 4E_g + 2A_{1g}$ %and the six Raman active phonon modes can be labeled using $D_{3d}$ point group. 
       Fig.~\ref{Fig_RS} shows the observed Raman spectrum for YbI$_3$ single crystal. The whole spectrum could be nicely fitted using a total of six Lorentzian lineshapes corresponding the six expected Raman active phonon modes. The modes are labeled from $\rm P_1$ to $\rm P_6$ in the increasing order of frequency as $\rm \Delta \omega = 38.4~(P_1), 56.4~(P_2), 69.4~(P_3), 94.0~(P_4), 118.5~(P_5)$, and $\rm 142.5~(P_6)$~cm$^{-1}$. The modes P$_1$, P$_3$, P$_4$, and P$_6$ have the $\rm E_g$ symmetry, whereas symmetry of P$_2$ and P$_5$ is $\rm A_{1g}$. Additionally, four weak modes at $\rm 61.6~(M_1), 74.3~(M_2), 114.1~(M_3)$, and $\rm 137.5~(M_4)$ are also observed. The presence of weak extra modes in the Raman spectrum are not unusual. These are often assigned as infrared active modes rendered Raman active due to lowering of the local symmetry at some sites~\cite{https://doi.org/10.1002/jrs.1250140202}. In some cases, the extra modes are assigned to second-order Raman scattering~\cite{Kuo2016}. The observed spectrum (including some of the weak extra modes) agrees well with that reported for the isostructural compound DyI$_3$~\cite{ChrissanthopoulosZissiPapatheodorou+2005+739+748}. The observed phonon modes for YbI$_3$ and DyI$_3$, and their corresponding symmetries in the $D_{3d}$ representation, are listed in Table S4.(see Supplemental Material)~\cite{supplemental}.  

       \label{Fig_RS}
        \begin{figure}[hbt]
		\centering
		\includegraphics[width = \columnwidth]{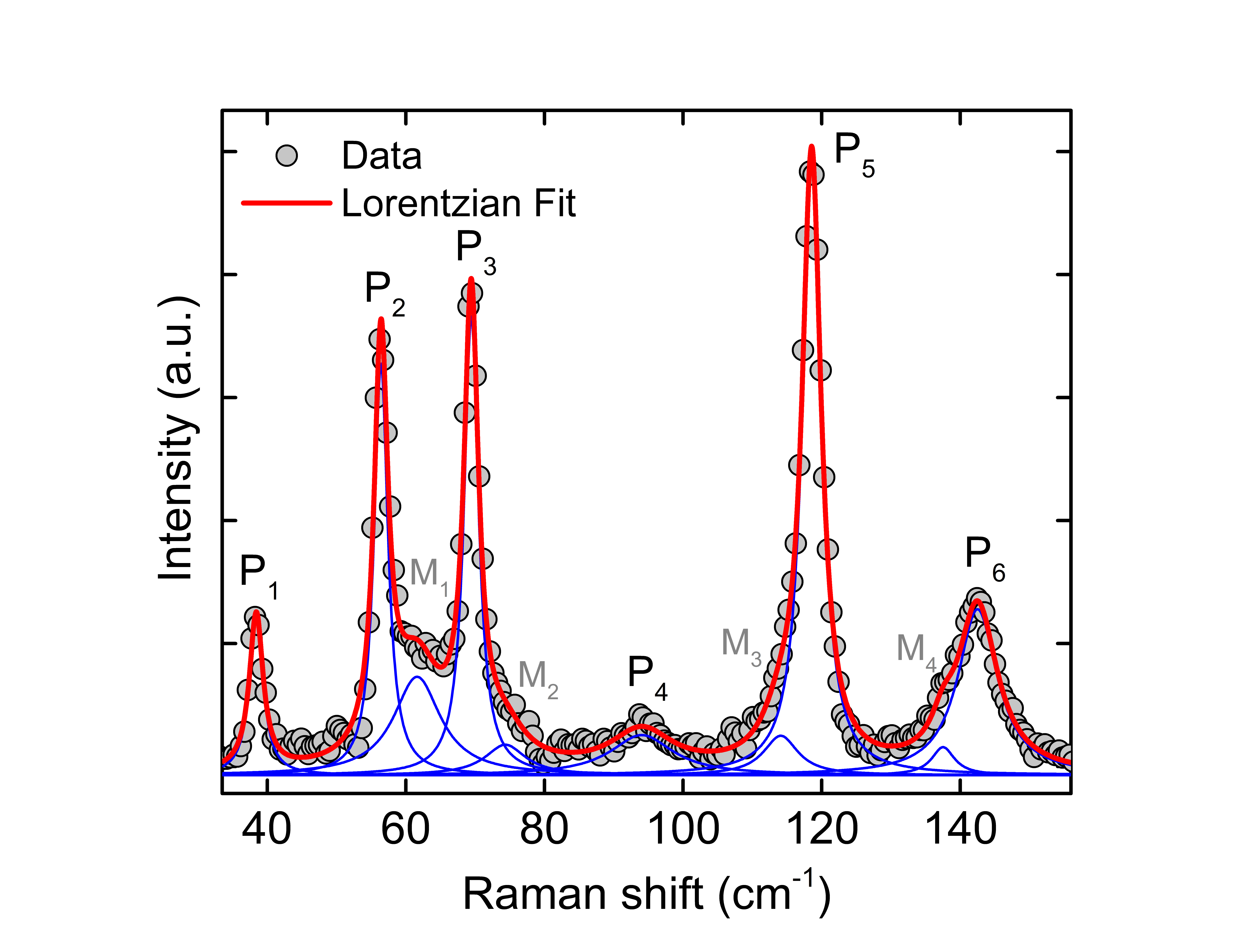}
		\caption {Raman spectrum of \Yb~ single crystal. The six Raman active phonon modes are labelled from $P_1$ to $P_6$. }
		%\label{Fig_RS}
	\end{figure}
	    
%\begin{table}[!]
%\caption{\label{Table_RS}%
%Experimental Peak position for \Yb~ and %DyI$_3$ at room temperature}
%\centering
%\begin{ruledtabular}
%\begin{tabular}{c c c c}
%Label & Positions($cm^{-1}$) & %Positions($cm^{-1}$) & Mode($D_{3d}$)  \\
% &  YbI$_3$ (Present work) & %DyI$_3$~\cite{ChrissanthopoulosZissiPapatheodo%rou+2005+739+748} &   
%\hline\\[0.5ex]
% $P_1$ & 38.41 & 39 & $E_g$ \\
% $P_2$ & 56.42 & 55 & $A_{1g}$ \\
% $P_3$ & 69.43 & 71 & $E_g$ \\
% $P_4$ & 94.04 & 94 & $E_g$ \\
% $P_5$ & 118.56 & 120 & $A_{1g}$ \\
% $P_6$ & 142.51 & 146 & $E_g$ \\
% $M_1$ & 61.63 & - & - \\
% $M_2$ & 74.35 & - & - \\
% $M_3$ & 114.10 & - & - \\
% $M_4$ & 137.53 & 139 & - \\
%\end{tabular}
%\end{ruledtabular}
%\end{table}
    
	\subsection{Heat Capacity}
        \label{HC}
        \begin{figure}
		\centering
		\includegraphics[width= 0.88\columnwidth]{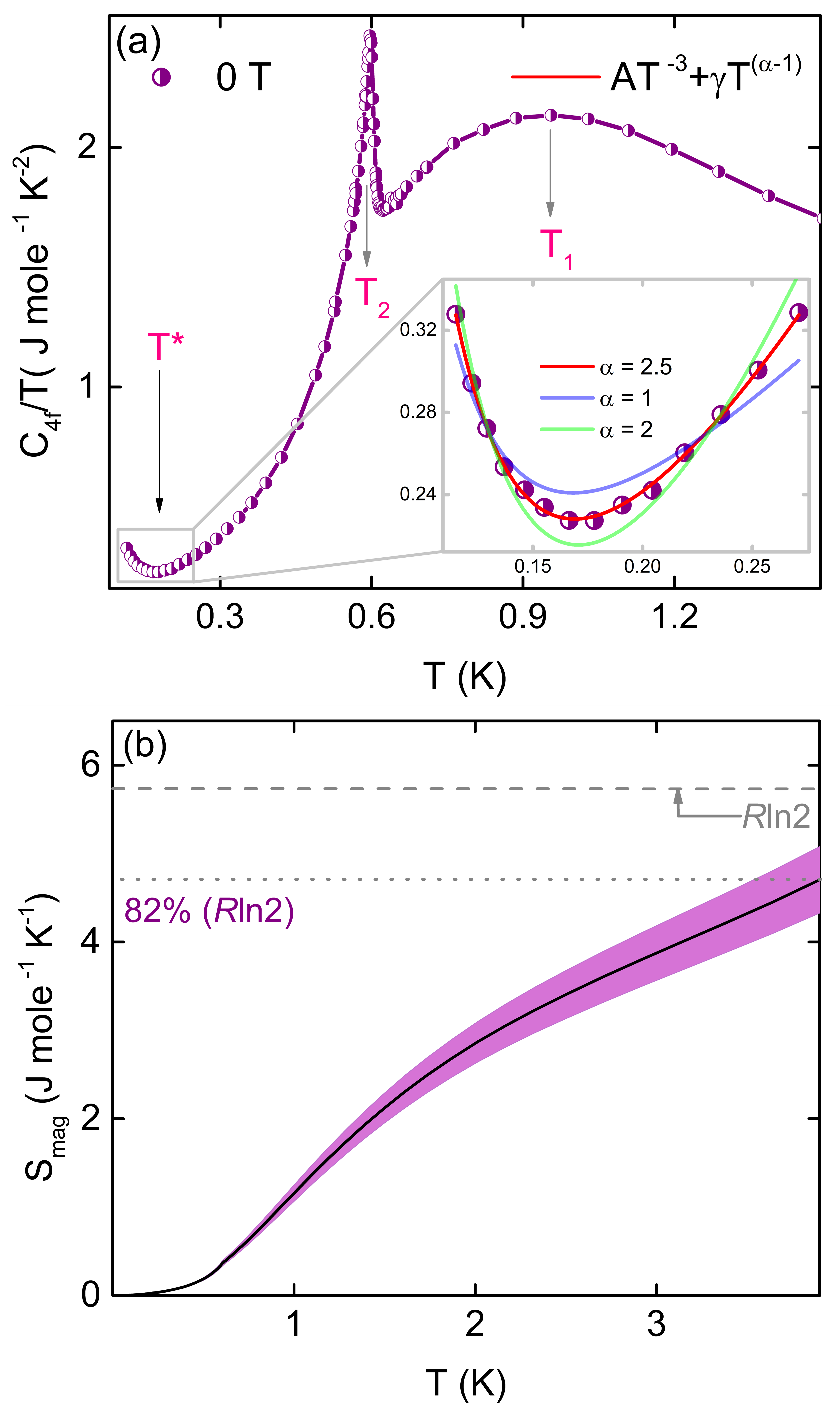}
		\caption {The temperature variation of heat capacity ($\rm C_p$) and magnetic entropy released ($\rm S_{mag}$): (a) $C_p/T$ is shown as a function of temperature. Inset: solid red line is a fit to the low temperature experimental data for $\alpha = 2.5$. The fit using $\alpha = 1, 2$ is also shown for reference (see text for details).   (b) $\rm S_{mag}$ is shown as a function of temperature. The purple shaded region indicates the error bars.}
		\label{Fig_HC}
	\end{figure}
        
        Fig.~\ref{Fig_HC}(a) shows the heat capacity $\rm C_p$ plotted as $\rm C_p/T$ as a function of temperature from $\rm 0.1~K \leq T \leq 1.5~K$. The heat capacity 
        is an additive quantity, comprising various contributions, and can be expressed as: $\rm C_p(T)=C_{mag}+C_{latt}+C_{nuc}$, where $\rm C_{mag}$ is contribution to heat capacity arising from the 4\textit{f} electrons of Yb$^{3+}$, $\rm C_{nuc}$ is %due to the nuclear electric quadrupole or due to 
        the nuclear Zeeman contribution, and  $\rm C_{latt}$ is the lattice contribution. In the temperature range of measurement, the lattice contribution due to phonons becomes extremely small and can therefore be neglected. %Consequently, Fig.~\ref{Fig_HC}(a) shows the heat capacity contributions from $C_{mag}$ and $C_{Nuc}$. 
        As shown in Fig.~\ref{Fig_HC}(a), the $\rm C_p/T$ exhibits three distinct features at temperatures T$_1$, T$_2$, and T$^\ast$. The broad anomaly around 0.95 K, labeled as $T_1$, is attributed to the short-range ordering of the Yb$^{3+}$ moments, whereas the sharp $\lambda$-type anomaly at 0.6 K, labeled as $T_2$, has been tentatively attributed to their long-range ordering. The upturn below T$^\ast$ is due to nuclear Zeeman splitting caused by the internal field generated by the ordering of Yb moments~\cite{Hallas_Yb_pyro}.   
               
        Here, it would be instructive to bring a comparison with much more extensively studied YbCl$_3$. In case of YbCl$_3$, a similar broad peak attributed to short-range ordering appears at 1.2~K, followed by weaker anomalies at 0.6~K and 0.4~K upon further cooling~\cite{PhysRevB.102.014427}. 
        While the 0.6~K anomaly is linked to the long-range ordering of the Yb moments, the the 0.4 K anomaly is sample-dependent and has been attributed to the presence of defects/imperfections or impurities. In contrast, in the \Yb~, the peak corresponding to the long-range ordering at T$_2$ = 0.6~K is sharp and well-defined, while the 0.4~K anomaly could not be detected in YbI$_3$. %The upturn below 0.3 K labelled as $T^*$ is due to nuclear Zeeman splitting caused by the internal field generated by the ordering of Yb moments [takagi et al]. 
        
        We fitted the heat capacity data of \Yb~ below 0.3 K using the expression: $\rm C_p=AT^{-2}+\gamma T^{\alpha}$, where the first term in the equation is the high-temperature approximation of the two-level Schottky equation~\cite{Steppke_Yb_HF_2010}, and the second term is written based on the assumption that $\rm C_{mag}$ follows a certain power-law behavior in the ordered state. The data fits well as shown with a red solid curve in the inset of Fig.~\ref{Fig_HC}(a). The values obtained for the parameters A, $\gamma$ and $\alpha$ from the fitting are: $\rm A=3.67\times10^{-4}~J~ mole^{-1}$~K, $\rm \gamma = 2.2~J~mole^{-1}~K^{-3.5}$, and $\alpha = 2.5$. For YbCl$_3$, the exponent $\alpha$ is $1.5$~\cite{matsumoto2024quantum}.The fit using the $\alpha = 1, 2$ is shown in the inset of Fig.~\ref{Fig_HC}(a) using blue and green solid curves, respectively. The fitting clearly indicates that the value of exponent $\alpha$ is 2.5 in case of YbI$_3$. In the magnetically ordered state of traditional antiferromagnets, the exponent $\alpha$ is expected to be 3, whereas in the case of layered antiferromagnets the expected value of $\alpha$ should be 2. On the other hand, $\alpha \approx ~ 1$, indicating a linear temperature dependence of low-temperature heat capacity conventionally holds for the spinons in gapless quantum spin liquids; whereas a more unconventional quadratic temperature dependence ($\alpha \approx 2$) is expected for a gapless Dirac fermions~\cite{PhysRevLett.98.117205}.The value of $\alpha \approx 2.5$ in the present case of YbI$_3$ is possibly an indication of an unusual ground state. Further, experiments in the presence of applied magnetic field, currently not accessible to us, will be illuminating in understanding the ground state of \Yb~ accurately.  
        
        Fig.~\ref{Fig_HC}(b) shows the magnetic entropy released as a function of temperature calculated using $\rm S_{mag} =\int_{0}^{T_0}C_{mag}(T)/TdT$. We extracted $\rm C_{mag}$ by subtracting the nuclear contribution from the total heat capacity. As mentioned in the beginning (Sec. \ref{Exp}), the addenda accounting for the contribution from the Apiezon N grease used for thermal contact and for covering the same to prevent it from decomposing in the process of loading for low-temperature measurements has not been subtracted. However, at low-temperatures, where $\rm C_{mag}$ is overwhelmingly large, we do not expect the Apiezon N contribution to be of much significance. The temperature variation of $\rm S_{mag}$ is shown in Fig.~\ref{Fig_HC}(b). Above 0.5~K, $\rm S_{mag}$ rises sharply, tapers a bit around 2~K, but do not show any signs of saturation up to 4~K. At 4~K, the value approaches 82\% of Rln2. Since the measurements were only conducted up to 4 K, the exact temperature at which the magnetic entropy plateaus at Rln2 is not known. However, previous reports on YbCl$_3$~\cite{PhysRevB.102.014427} suggest that this plateau will possibly occur at a slightly higher temperature ( close to 10 K). A value of $\rm S_{mag}~\approx$~Rln2, shows that the crystal field split ground state of Yb$^{3+}$ in \Yb~ is a Kramer's doublet. To establish this further, we investigated the magnetic behavior of our \Yb~ crystals, disccussed next. 
               
        %As expected for a Spin 1/2 system, the total entropy release is close to $Rln(2S+1) \approx 5.76~J\cdot mole^{-1} \cdot K^{-1}$.

	\subsection{Magnetization}
        \label{Mag}

        \begin{figure}[!]
		\centering
		\includegraphics[width= 0.92\columnwidth]{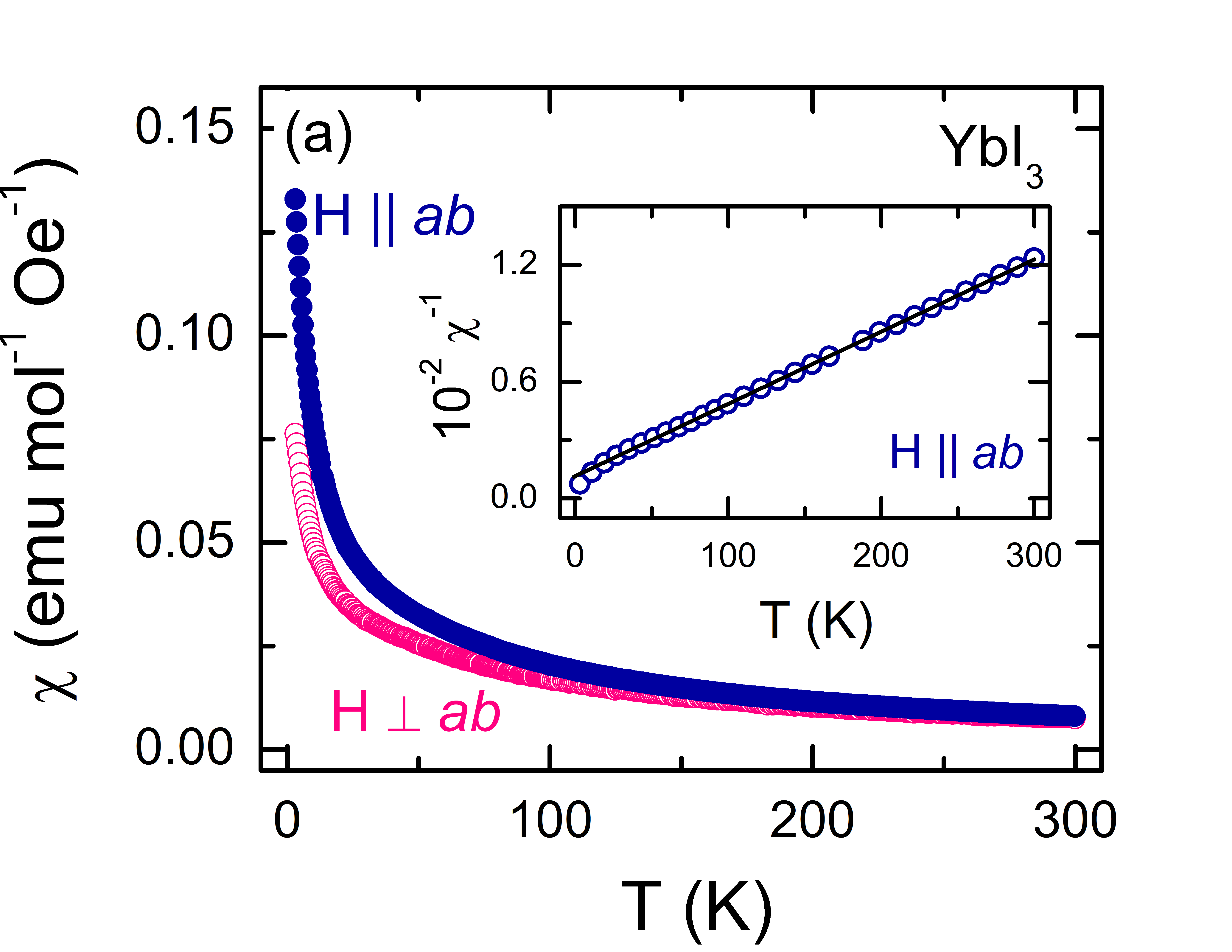}\\
        \includegraphics[width= 0.92\columnwidth]{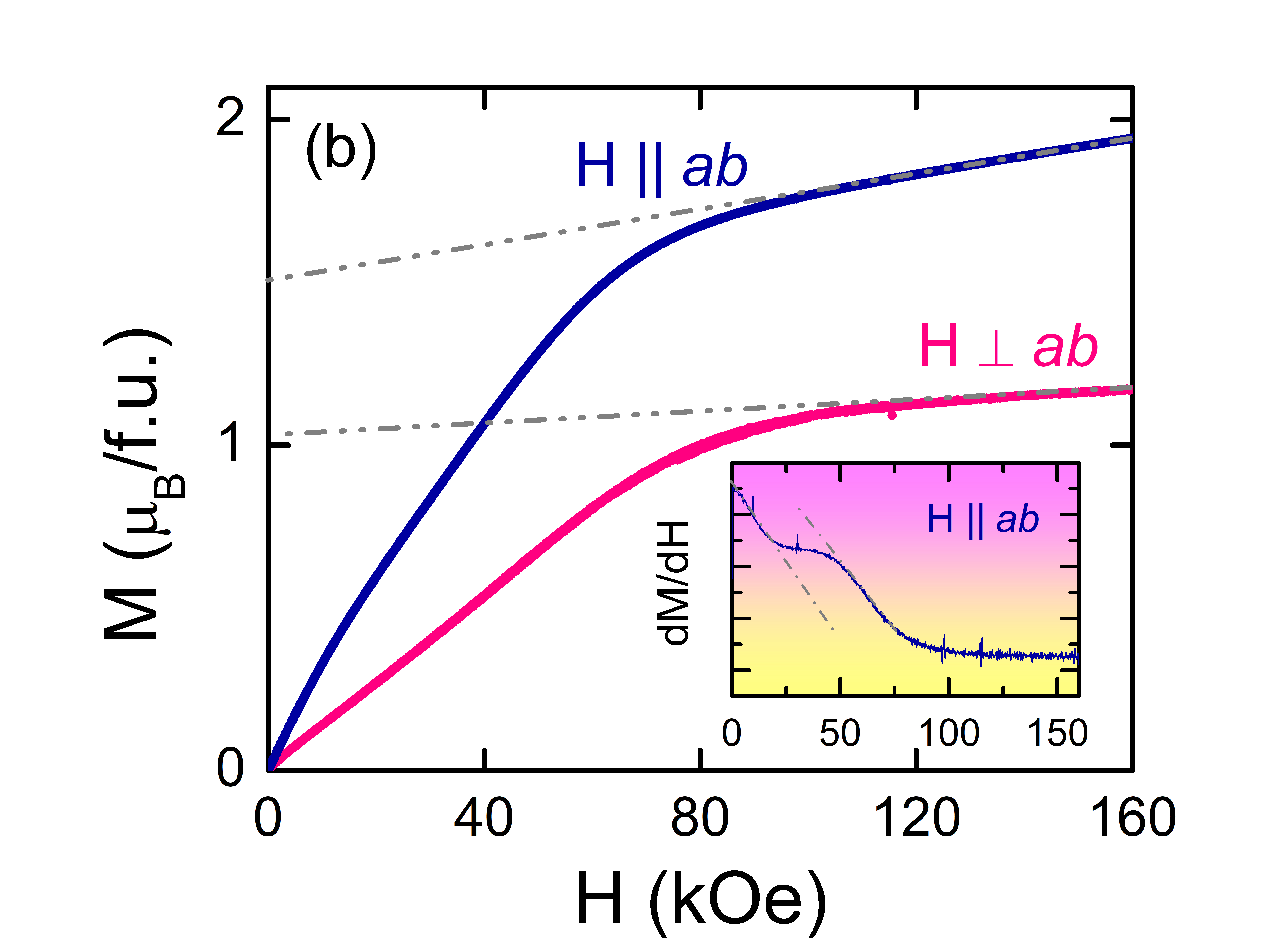}
  		\caption {The magnetic susceptibility ($\chi$) and isothermal magnetization M(H) of \Yb~: (a) Temperature variation of $\rm \chi_\parallel$ and $\rm \chi_\perp$ measured under an applied field magnetic field of 10~KOe. Inset: $\rm \chi^{-1}(T)$ plotted as a function of temperature for $H \parallel ab$. The solid line through the data points in the inset is the modified Curie-Weiss fit (see text for details). (b) M(H) at 2 K for $H \perp ab$ and $H \parallel ab$. The dashed lines represent linear extrapolation of the high-field data to $\rm H = 0$. Inset: derivative dM/dH plot.}
		\label{Fig_MT}
	\end{figure}

  %\begin{figure}
		%\centering
		%\includegraphics[width= \columnwidth]{Figure_6_MH.png}
		%\caption {The field variation of isothermal magnetization (M) of \Yb~: (a) M(H) at 2 K for $H \parallel ab$. The solid line is a linear extrapolation of the high-field data to $\rm H = 0$. Inset: M(H) is shown in the range $ -10~T \leq H\leq 10~T$ (left y-axis), and the its derivative. dM/dH (right y-axis).(b) M(H) at 2~K for $H \perp ab$. The solid line is a linear extrapolation of the high-field data to H = 0. Inset: a zoomed-in view of the region where the M(H) bends. \textcolor{red}{THIS FIGURE SHOULD BE REDONE}}
		%\label{Fig_MH}
	%\end{figure}
 
        Magnetic susceptibility ($\chi$) is measured by applying a magnetic field of 10 kOe, in the parallel ($\chi_{\parallel}$) (in-plane) and perpendicular ($\chi_{\perp}(T)$) (out-of-plane) orientations. The temperatures variation of in-plane and out-of-plane susceptibilities are shown in Fig.~\ref{Fig_MT}(a). For the out-of-plane orientation, the zero-field-cooled (ZFC) and field-cooled (FC) curves overlap over the whole temperature range, but in the in-plane susceptibility, a very weak splitting is observed below approximately 150$~K$ (not shown).%, as shown in the inset of Fig.~\ref{Fig_MT}(a). 
            
            The high temperature $\rm 1/{\chi}$ data ($150 \leq T \leq 300$) are fitted using a modified Curie-Weiss (CW) expression: $\rm {\chi} = {\chi_0 + C/(T - \Theta_{CW})}$, where C is the Curie-constant from which the value of the effective magnetic moment ($\rm \mu_{eff}$) can be derived using $\rm \mu_{eff} = \sqrt{8C}$; $\rm \Theta_{CW}$ is the Weiss temperature, and $\rm \chi_0$ is a small temperature independent term that usually arise due to core diamagnetism and/or van Vleck type paramagentism. 
            
        The values of the fitting parameters for the in-plane and out-of-plane orientation are as follows: $\rm \chi_0^{\perp} = 3.8\times10^{-4}~emu~mole^{-1}~Oe^{-1}$%($ \rm \chi_0^{\perp} = 1.9\times10^{-4}~emu~mole^{-1}~Oe^{-1}$)
            , $\rm \Theta_{cw}^\perp = -55~K$% ($\rm \Theta_{cw}^{\perp} = -24~K$)
            , and $\rm \mu_{eff}^{\perp} = 4.53~\mu_B/Yb^{3+}$% ($\rm \mu_{eff}^{\perp} = 4.54~\mu_B/Yb^{3+}$
            ; the corresponding values in the in-plane orientation are: $ \rm \chi_0^{\parallel} = 1.9\times10^{-4}~emu~mole^{-1}~Oe^{-1}$, $\rm \Theta_{cw}^{\parallel} = -24~K$, and $\rm \mu_{eff}^{\parallel} = 4.53~\mu_B/Yb^{3+}$. 
            In both cases, the value of $\rm \mu_{eff}$ agrees well with the calculated value of $\rm \mu_{eff} = g_J\sqrt{J(J+1)} = 4.53 \mu_B$ for Yb$^{3+}$ ($4f^{13}$: $\rm L = 3, S = 1/2$ , and $\rm J = 7/2$). The large values of $\Theta_{CW}$ for both the orientations are unrealistic, which may be due to the large crystal field splitting of the J = 7/2 manifold. %This may also be the reason for somewhat higher values of $\rm \chi_0$. 
            In the presence of an octahedral crystal field environment of surrounding I$^-$ ions, the eight-fold degenerate, $\rm J = 7/2$ manifold in a free Yb$^{3+}$ ion will split into four Kramers doublet. As a reference, we note that in YbCl$_3$, the four Kramer's doublets are located at energies, 0, 243, 371, and 456~K. Clearly, in the range over which the fitting is performed, the van Vleck contribution from the higher-lying levels will be small but not insignificant. %A slight change in slope of $\rm 1/{\chi}$ versus T plot near 60~K [Fig.~\ref{Fig_MT}(b)] is likely a manifestation of the crystal field splitting. 

        \begin{table*}[t],
		\setlength{\tabcolsep}{12pt}
		\caption{Summary of fitting parameters $\chi_0$ ($\rm emu~mole^{-1}~Oe^{-1}$), C ($\rm emu~mole^{-1}~Oe^{-1}~K$), $\rm \mu_{eff}$ ($\rm \mu_B$), anisoptropic Lande g-factor and $\rm \Theta_{cw}$ (K) obtained by fitting the magnetic susceptibility data for in-plane ($H \parallel ab$) and out-of-plane ($H \perp ab$) orientations}
		\label{Table_CWF}
		\centering
		\begin{ruledtabular}
        \begin{tabular}{c c c c c c c c}
				Orientation & Temperature range (K) & $\chi_0$ & C  & $\rm \mu_{eff}$  & g-factor & $\rm \Theta_{CW}$ & Ref.\\[0.5ex]
				\hline\\[0.5ex]
				$H \perp ab$ & $150-300$ & $3.82\times10^{-4}$ & $2.57$  & $4.53$ & $-$ & $-55$ &  \\ [0.5ex]
				  & $4-12$ & $0~^a $ & $0.98$ & $2.80$ & $3.2$ & $-10$ & This work\\ [1.0ex]
                    & $4-12$ & $5.5\times10^{-3}~^b$ & $0.82$ & $2.56$ & $3.0$ & $-8$ & \\ [1.0ex]
                    & $^{\textcolor{blue}{c}}$ $3-15$ & $0$ & $1.12$  & $3.0(1)$ & $3.5(1)$ & $-9$ & ~\cite{PhysRevB.102.014427}\\ [0.5ex]
				$H \parallel ab$ & $150-300$ & $1.95\times10^{-4}$ & $2.57$  & $4.53$ & $-$ & $-24$ & \\ [0.5ex]
				  & $4-12$ & $\sim 0~^a$ & $1.32$ & $3.24$ & $3.8$ & $-7$ & This work\\ 
                    & $4-12$ & $16\times10^{-3}$~$^b$& $0.93$ & $2.73$ & $3.2$ & $-5$ &  \\ 
                    & $^{\textcolor{blue}{c}}$ $3-15$ & $0$ & $1.20$  & $3.1(1)$ & $3.6(1)$ & $-6$ & ~\cite{PhysRevB.102.014427}\\ [0.5ex]
      
		\end{tabular}
        \end{ruledtabular}
        \footnotesize{$^a$~when $\chi_0$~is fixed as 0; $^b$ When $\chi_0$ is fixed to the value obtained from the slope of the linear-section of the high field isothermal magnetization plots. $^{\textcolor{blue}{c}}$~YbCl$_3$
        }
            \end{table*}
            
        In such scenarios where the crystal field splitting is large, it is more instructive to fit the low-temperature data to estimate the magnetic moment in the lowest Kramers doublet.           
            %using the modified Curie weiss law with $\chi_0 = \chi^{VV}$ obtained from the M(H) isotherm explained later in the text. The fit yields $\chi_{VV}^{\parallel} = 5.558\times10^{-3} emu\cdot mole^{-1}\cdot Oe^{-1}$ and $\chi_{VV}^{\perp} = 16.20\times10^{-3} emu\cdot mole^{-1}\cdot Oe^{-1}$ , 
            We fitted the data in the temperature range from 4 K to 12 K using the modified Curie-Weiss expression. The values of the fitting parameters thus obtained are listed in Table. \ref{Table_CWF}. For both the orientations, if we fix $\rm \chi_0 = 0$, we get $\rm \mu_{eff}^\perp~=~$2.8~$\rm \mu_B/Yb^{3+}$, $\rm \mu_{eff}^\parallel = 3.24$~$\rm \mu_B/Yb^{3+}$; and, $\rm \Theta_{CW}^\perp = -10$~K, $\rm \Theta_{CW}^\parallel = -7~K$. In YbCl$_3$, the corresponding values are~\cite{PhysRevB.102.014427}: $\rm \mu_{eff}^\perp~=~$3.0~$\rm \mu_B/Yb^{3+}$, $\rm \mu_{eff}^\parallel = 3.1~\mu_B/Yb^{3+}$; and, $\rm \Theta_{CW}^\perp = -9$~K, $\rm \Theta_{CW}^\parallel = -6$~K. The two sets of values are in good agreement with each other, suggesting some degree of similarity between the two systems. %Clearly, the magnetic moment in the crystal field split ground state of Yb$^{3+}$ in \Yb~ is smaller than in YbCl$_3$. The values of Weiss temperatures for the two compounds is comparable. 
            In both cases, the value of $\Theta_{CW}$ is somewhat higher than expected as we do not expect the exchange interaction to be of the order of 10 K for the reason that the 4\textit{f}-electrons in rare-earths are highly sequestered, shielded by \textit{5d} and \textit{6s} electrons, which prevents them from making an appreciable overlap with the ligand \textit{p} orbitals. For example, in the Yb-based pyrochlore oxides, the exchange interaction is of the order of 1~K. The other problem with these values is that the estimated Weiss temperature is comparable from the two orientations, which appears improbable given the layered nature of these materials, which suggests that the interaction within the plane should be much stronger than out-of-plane. In YbCl$_3$, the in-plane nearest-neighbor coupling strength is $\rm \Lambda_\parallel~\approx~5$~K and $\rm \Lambda_\perp/\Lambda_\parallel \approx 10^{-5}$~\cite{sala2021van}. The exact  reason for the high Weiss temperature in the low-temperature Curie-Weiss fits is not yet clear. Fixing $\chi_0$ to some high value ($\rm \sim~10^{-4}-~10^{-3}~emu~mole^{-1}~Oe^{-1}$)~did not decrease the Weiss temperature significantly either. We believe that the short-range magnetic correlations, that persists up to very high temperatures, may have contributed to the high value of $\rm \Theta_{CW}$. Strong spin fluctuations or short-range correlations are intrinsic to the layered (or quasi-2D) magnets, which is reflected in the broad peak in the heat capacity and a nearly linearly increasing $\rm S_{mag}$ up to temperatures as high as $\sim$~7T$_N$ (here T$_N$ = T$_2$ corresponds to the sharp anomaly in $\rm C_p$). %Same observation is valid for YbI$_3$. 
            The value of anisotropic g-factor in \Yb~ is estimated to be $\rm g_J^\perp = 3.2$ and $\rm g_J^\parallel = 3.8$, assuming $\rm J_{eff} = 1/2$. %which suggests a nearly Heisenberg-like behavior of the Yb moments.
            In comparison, from a similar analysis the anisotropic g-factor in YbCl$_3$ is $\rm g_J^\perp = 3.5$ and $\rm g_J^\parallel = 3.6$~\cite{PhysRevB.102.014427}. Slightly smaller values for the g-factors are obtained by taking $\chi_0$ values from the isothermal magnetization, as discussed below.
                      
            %The negative sign of $\Theta_{CW}$ for both orientations is negative, indicating that the magnetic interactions are antiferromagnetic both 
            %$\Theta_{cw}^{\parallel} = -8.4 K$ and $\Theta_{cw}^{\perp} = -4.9 K$, $\rm \mu_{eff}^{\parallel} = 2.56\mu_B $ and $\rm \mu_{eff}^{\perp} = 2.733\mu_B $, the g-value obtained using $\rm \mu_{eff} = g_J\sqrt{J(J+1)}$ with $J_{eff} = 1/2$ are $g^{\parallel} = 2.96$ and $g^{\perp} = 3.16$. The negative values of Curie Weiss temperature indicates the antiferromagnetic exchange interactions between the spins in the ab plane and along the c - direction similar to YbCl$_3$~\cite{PhysRevB.102.014427}. According to the obtained values of the g - factor, the spins exhibit a Heisenberg-like nature without any noticeable anisotropy similar to the case of YbCl$_3$~\cite{PhysRevB.102.014427}. The values obtained for high-temperature and low-temperature fit for $H \perp ab$ and $H \parallel ab$ direction are summarized in Table \ref{Table_CWF}.       
            
        Isothermal magnetization, M(H) for out-of-plane ($H \perp ab$) and in-plane ($H \parallel ab$) orientations is shown in Fig.~\ref{Fig_MT} (b). In the out-of-plane orientation, the magnetization nearly saturates above a field of $H \approx 120$~kOe. The small slope that still persists can be attributed to the van Vleck term ($\chi_0$), which can be calculated by fitting the high-field data to an expression of the $\rm M=M_s^\perp$ + $\chi_0^\perp H$, where M$_s^\perp$ is the saturation magnetization in the out-plane orientation. The fit, which is shown in the Fig.~\ref{Fig_MT}(b) using a dashed line (shown extrapolated to H = 0), gives $\rm M_s^\perp \approx 1 ~\mu_B/Yb^{3+}$ and $\chi_0^\perp = 5.6\times 10^{-3}$~emu~mol$^{-1}$~Oe$^{-1}$. Using these values of $\chi_0$ for the orientations (instead of fixing them to zero, as done in the previous paragraph), we get $\rm g_J^\perp = 3.2$ and $\rm g_J^\parallel = 3.0$ (see Table.~\ref{Table_CWF}). Combining the two results, we can conclude that the approximate $\rm g_J$ values for the two orientations are: $\rm g_J^\perp \approx 3.0 \pm 0.2$ (out-of-plane), and $\rm g_J^\parallel \approx 3.4 \pm 0.3$ (in-plane).   
            
       For the in-plane M(H) a more intriguing field variation is observed, as depicted using the derivative ($d$M/$dH$) plot in the inset of Fig.~\ref{Fig_MT}(b). With increasing field, $d$M/$dH$ exhibits a plateau in the field ranging form 20~kOe to 40~kOe, suggesting a field induced transition. In the high-field region $d$M/$dH$ becomes almost flat and very small. Analyzing the high-field region as before yields: $\rm M_s^\parallel \approx 1.5 ~\mu_B/Yb^{3+}$ and $\chi_0^\parallel = 16\times 10^{-3}$~emu~mol$^{-1}$~Oe$^{-1}$.
            Using, the g-factor obtained for the two orientations from the susceptibility data, a back calculation of the saturation moment suggests $\rm J_{eff} \leq 1/2$.    

 \subsection{YbI$_3$ versus YbCl$_3$}
 
        To begin with, the crystal structures of YbI$_3$ and YbCl$_3$ are distinct from one another. YbI$_3$ exhibits trigonal symmetry, while YbCl$_3$ displays monoclinic symmetry~\cite{PhysRevB.100.180406}. This primary difference is due to the variation in the stacking sequence of YbX$_3$ layers along the \textit{c}-axis. In YbI$_3$, three layers form a repeating unit, resulting in nearest neighbor Yb-Yb distance of 6.803~\AA~along the \textit{c}-axis, as shown in Fig.~\ref{Fig_CS}(c). In contrast, YbCl$_3$ consists of a single layer repeated along the \textit{c}-axis, leading to nearest neighbors at 6.313~\AA~in the layers above and below the middle layer, as shown in Fig.~\ref{Fig_CS}(d). Beyond the global symmetry, the local symmetry within a single YbX$_3$ layer in the \textit{ab}-plane also differs slightly for the two compound, which may be crucial for understanding the magnetic behavior of these materials. In YbCl$_3$, the Yb-Yb honeycomb bond distances are 3.886~{\AA} and 3.864~{\AA}; accordingly, the bond angles are 96.12$^\circ$ and 96.73$^\circ$, respectively~\cite{PhysRevB.100.180406}. Conversely, YbI$_3$ features a uniform bond distance of 4.226(1)~{\AA} and a bond angle of 91.76(3)$^\circ$, resulting in an isotropic honeycomb lattice with equal bond distances and angles. Although the distortion in YbCl$_3$ is very small, it nevertheless causes the breaking of C$_3$ symmetry, which is preserved in the case of YbI$_3$. 
 
        Heat capacity measurements reveal two distinct transitions in both compounds. In YbI$_3$, short-range ordering occurs at 0.95~K, followed by long-range ordering at 0.6~K. Similarly, in YbCl$_3$, short-range ordering occurs at 1.2~K, followed by long-range ordering at 0.6~K. The entropy release for both compounds approaches Rln2, indicating a \J~= 1/2 Kramers doublet ground state. Regarding differences, the transition at 0.6~K attributed to the long-range ordering in YbI$_3$ is rather strong, while it is significantly weaker in YbCl$_3$. On the other hand, the broad hump due to short-range ordering is much more pronounced in YbCl$_3$ compared to YbI$_3$. If this broad hump is due to 2D spin correlations, as has been argued in the literature~\cite{sala2021van}, we expect it to be more pronounced in \Yb~, where the interlayer separation is larger. This aspects should be looked into more carefully in the future studies. Another notable difference in the heat capacity of two compounds is that in YbCl$_3$, a weak sample dependent anomaly is reported at 0.4~K, which is absent in YbI$_3$.

        Magnetic behavior of the two compounds above 2 K is qualitatively similar. Further low temperature measurements on \Yb~ might prove useful in this context. The saturation moments $\rm M_s^\parallel \approx 1.5 ~\mu_B/Yb^{3+}$ and $\rm M_s^\perp \approx 1 ~\mu_B/Yb^{3+}$ is nearly the same for both the compounds. The differences may crop-up at lower temperatures, hence, magnetic measurements below 2~K would be interesting.  %Magnetic susceptibility measurements reveal slightly greater anisotropy in YbI$_3$ compared to YbCl$_3$, as corroborated by the anisotropic g-factor values summarized in Table.~\ref{Table_CWF}

        In conclusion, YbI$_3$ possesses a more symmetric structure compared to YbCl$_3$, with C$_3$ symmetry preserved, similar to $\alpha$-RuCl$_3$, whereas YbCl$_3$ exhibits broken C$_3$ symmetry. The nearest neighbor Yb$-$Yb separation along the \textit{c}-axis is found to be larger in YbI$_3$ compared to YbCl$_3$. Heat capacity measurements show a similar two-step ordering for both compounds: initial short-range ordering followed by long-range antiferromagnetic ordering. Any further additional anomaly due to poor quality of the sample is absent in YbI$_3$. In the paramagnetic state (T $>$ 2 K), the magnetic behavior of YbI$_3$ resembles that of YbCl$_3$. %with slightly increased anisotropy as indicated by values obtained from magnetic susceptibility measurements. To precisely determine the spin anisotropy in YbI$_3$, more detailed studies using polarized neutron diffraction or inelastic neutron scattering are necessary.
 
 %Crystal structure:\\ 
 %Magnetic behavior: \\
 %Heat capacity: \\
 %A paragraph, summarizing this subsection}
 
 \section{Summary \& Conclusions}
            To conclude, we have successfully grown high-quality single crystals of YbI$_3$ using a vapor transport reaction, taking elemental Yb and I as the starting precursors. %The quality of these crystals was verified using various structural characterization tools, including, x-ray diffraction on the \textit{ab}-plane, SEM-EDX analysis, and single crystal x-ray diffraction. We show that YbI$_3$ crystallizes with $R\Bar{3}$ symmetry, similar to the low-temperature phase of $\alpha-$RuCl$_3$, with Yb forming regular hexagonal rings in the honeycomb layer. The symmetry is further validated by Raman spectroscopy at room temperature, where Six Raman modes at $\rm \Delta \omega = 38.4, 56.4, 69.4, 94.0, 118.2,~and~142.5$~cm$^{-1}$ are observed consistent with the symmetry derived from the single crystal x-ray diffraction. 
            The crystal field split ground state of Yb$^{3+}$ in YbI$_3$~is a well-isolated Kramer's doublet with  $\rm g_J^\perp \approx 3.0 \pm 0.2$ (out-of-plane), and $\rm g_J^\parallel \approx 3.4 \pm 0.3$ (in-plane). The in-plane isothermal magnetization shows weak a field induced transition, the indicating presence of slight anisotropy even within the \textit{ab}-plane.      
                       
            Heat capacity show a two-step magnetic transition with a broad peak T$_1$ = 0.95 K due to short-range correlations between the Yb moments, followed by a sharp peak at 0.6~K, signaling transition into a long-range magnetically ordered state. No additional anomaly, except an uptrun below about 0.25~K due to nuclear Schottky effect, could be seen.%, differentiating YbI$_3$ from YbCl$_3$, where a third, sample-dependent, peak has been reported.
            %What else distinguishes YbI$_3$ form YbCl$_3$ is the relative magnitudes of T$_1$ and T$_2$ peaks. In YbCl$_3$, the broad peak at T$_2$ (= 1.2~K) is most dominant, whereas in YbI$_3$, it is the sharp peak at T$_1$.   
            The magnetic entropy, %estimated after subtracting the nuclear contribution, is found to grow rapidly above T = 1 K, 
            reaches almost 80\% of Rln2 near 4~K ( highest temperature in our experiment), consistent with the Kramer's doublet ground state. 
            %suggests short-range spin correlations, while the second, at T$_2$ = 0.6 K, indicates long-range antiferromagnetic ordering. The low-temperature upturn, marked as T$^*$, was analyzed using a nuclear Schottky anomaly and a power law, 
            The value of critical exponent $\alpha$ of 2.5 from the heat capacity data %lies between the 3D Heisenberg antiferromagnetic ($\alpha$ = 3) and gapless Dirac fermions ($\alpha$ = 2), 
            suggests a possible unconventional ground state for YbI$_3$. To conclude, YbI$_3$ %emerges as a potential candidate for either a Kitaev quantum spin liquid or a 3D Heisenberg antiferromagnet on a 2D honeycomb lattice. Both are novel states of matter, 
            appears to harbor very interesting physics associated with \J~ = 1/2 on an ideal 2D honeycomb lattice. We believe that our preliminary study will trigger further experimental investigation to explore the spin Hamiltonian and exact ground state of this compound.

            %The magnetic susceptibility is weakly anisotropic with $\chi^\perp > \chi^\parallel$ over the whole temperature range. The effective moment per Yb$^{3+}$ from the high-temperature fit agrees well with the free-ion value of 4.54 $\mu_B$. At low temperatures, the moments are considerably reduced. The effective moment in the crystal field split lowest Kramer's doublet                      
            %Magnetization measurements revealed similar g-factors along two directions($H\parallel c$ and $H\perp c$), indicating Heisenberg-like spin behavior. 
        
	\label{SC}
	
	\section*{Acknowledgments}
The authors acknoweldge UGC-DAE-CRS user facility for magnetization and low-temperature specific heat measurements. S.R. acknowledges I-HUB, National Mission on Interdisciplinary CyberPhysical Systems (NM-ICPS) of the Department of Science and Technology, Government of India, for the postdoctoral fellowship. L.H. and S.S. thank I-HUB (NM-ICPS) for partial financial support. S.S. thanks financial assistance received under the Scheme for Transformational and Advanced Research in Sciences (STARS) by the Ministry of Education, Government of India (sanction order no. STARS/APR2019/PS/358/FS). N.P. thanks Council for Scientific and Industrial Research (CSIR) for Ph.D. fellowship. P.T. thanks DST INSPIRE for the Ph.D. fellowship.
 
	\bibliography{YbI_ref}
	\bibliographystyle{apsrev4-2.bst}

\end{document}